\documentclass[pra,aps,twocolumn,showpacs]{revtex4}
\usepackage[]{times,amsmath,epsfig,graphicx,color}
\definecolor{darkblue}{rgb}{0,0,0.75}
\definecolor{darkgreen}{rgb}{0.,0.7,0}
\definecolor{darkred}{rgb}{0.7,0,0}
\definecolor{note}{rgb}{0.25,0.0,0.75}
\usepackage[unicode, colorlinks, citecolor=darkblue, linkcolor=darkred, urlcolor=blue]{hyperref}
\allowdisplaybreaks
\begin{document}

\title{Optical Gravitational Wave Antenna with Increased Power Handling Capability}

\author{ Andrey B. Matsko$^{1}$, Mikhail V. Poplavskiy$^2$, Hiroaki Yamamoto$^3$, and Sergey P. Vyatchanin$^{2}$}

\affiliation{$^1$ OEwaves Inc., 465 North Halstead Street, Suite 140, Pasadena, CA 91107, USA}

\affiliation{$^2$  Faculty of Physics, Lomonosov Moscow State University, Moscow 119991, Russia}

\affiliation{$^3$ LIGO Laboratory, California Institute of Technology, MC~100-36, Pasadena, CA 91125, USA}

\begin{abstract}
Fundamental sensitivity of an optical interferometric gravitational wave detector increases with increase of the optical power which, in turn, limited because of the opto-mechanical parametric instabilities of the interferometer. We propose to optimize geometrical shape of the mirrors of the detector to reduce the diffraction-limited finesse of unessential optical modes of the interferometer resulting in increase of the threshold of the opto-mechanical instabilities and subsequent increase of the measurement sensitivity. Utilizing parameters of the LIGO interferometer we found that the proposed technique allows constructing a Fabry-Perot interferometer with round trip diffraction loss of the fundamental mode not exceeding $5$~ppm,  whereas the loss of the first dipole as well as the other high order modes exceed $1,000$~ppm and $8,000$~ppm, respectively. The optimization comes at the price of tighter tolerances on the mirror tilt stability, but does not result in a significant modification of the optical 
beam profile 
and does not require changes in the the gravity detector read-out system. The cavity with proposed mirrors is also stable with respect to the slight modification of the mirror shape.
\end{abstract}


\maketitle

\section{Introduction}

Gravitational wave astronomy inherently relies on high power resonant optical systems. The power of the probe light circulating in a cavity is the ultimate lever utilized to increase the sensitivity of a position measurement of a gravitational wave detector test masses carrying information about gravitational wave signals.  The projected continuous wave (cw) light power pushes 0.8~MW value in the second generation of gravitational wave detectors, such as Advanced LIGO (now in operation), Advanced VIRGO and KAGRA, planned to become operational in the next few months  \cite{aLIGO2013,aLIGO2014}. While this power value is by far lower if compared with the optical damage limit of the cavity mirrors, it is high enough to initiate various nonlinear processes resulting in depletion of the probe light and in generation of optical harmonics adding noise to the recorded signal and hindering the desirable sensitivity increase. Technical solution allowing suppressing the nonlinear interactions are needed to push the 
limits of gravitational wave 
astronomy and to widen the horizon of observable events associated with gravitational wave emission.

Resonant opto-mechanical oscillations  are expected to have the lowest power threshold if compared with the other nonlinear processes in the cavities, it may  cause  undesirable parametric instability (PI) \cite{01a1BrStVy,02a1BrStVy}. The PI occurs due to interaction of optical cavity modes and mechanical modes of the cavity mirrors. The photons of the probe light confined in a selected, usually fundamental, cavity mode pumped at frequency $\omega_p$ are parametrically converted to mechanical phonons of the cavity mirrors (having frequency $\Omega_m$) as well as lower frequency, or Stokes, photons emitted into high order optical modes having frequency $\omega_s\simeq \omega_p-\Omega_m$. The power threshold of PI is inversely proportional to the product of quality factors of the optical and mechanical modes participating in the process, so desirable reduction of the optical as well as mechanical attenuation results in undesirable reduction of the PI threshold.

The phenomenon of PI was studied and validated experimentally in a table top Fabry-Perot resonator \cite{14ArxivChen} as well as in whispering gallery mode resonators \cite{05Kippenberg, 09PRLmatsko, 12OEmatsko}. Recently PI was observed in full scaled Advanced LIGO interferometer \cite{15ArxivEvans} at relatively small circulating power $\sim 50$~kW as compared with $0.8$~MW planned in a Advanced LIGO.

Efficiency of PI depends on phase matching, comprising nonzero overlap integral and energy conservation, of the optical and mechanical modes. There is a significant probability that these conditions are always fulfilled in long-base gravitational wave detectors because of dense spectrum of optical modes of large cavities and dense spectrum of mechanical modes of large area cavity mirrors. Since the mirrors involved into the system are not identical, they have slightly different associated mechanical frequencies that can lead to PI.

Several techniques of reducing PI impact have been studied recently. They involve either braking the phase matching of the nonlinear process by changing frequency spectra of the modes participating in the IP process, or reducing PI efficiency by  damping nonessential modes.

For instance, one can move the opto-mechanical system out of resonance by controlling surface temperature of the mirrors \cite{05PRLzhao, 10CQGgrass, 07JOSABdegallaix}. This is possible since the optical ($\omega_p$ and $\omega_s$) and mechanical ($\Omega_m$) eigenfrequencies of the system depend on the mirror temperature $T$ in different ways, so the PI favorable condition $\omega_p(T)=\omega_s (T)+\Omega_m (T)$, ultimately breaks. The drawback of this technique is related to its lack of selectivity. All the modes of the optical cavity move at nearly the same pace, and while one pair of Stokes and mechanical modes comes out of the resonance, another pair comes in. This drawback can be partially suppressed by modulation of the temperature of the mirror surface.

Alternative stabilization method involves damping of mechanical modes either in a passive or an active way. It was found that introducing an annular strip at the rims of cavity mirrors reduces quality (Q-) factors of elastic modes \cite{09GQGgras, 10CQGgrass, 09CQGju}. However, this strip reduces Q-factor of the modes within the whole spectrum, including reception band of the antenna ($30\dots 500$~Hz). This is undesirable, since low mechanical attenuation at these frequencies is essential for achieving the desirable detection sensitivity.

Active electro-mechanical feedback allows reducing Q-factor of several particular elastic modes \cite{02a1BrVy,11PRAmiller}. The method is too selective to suppress all high-frequency modes in the entire bandwidth of interest ($50 \dots 200$~kHz) and, hence, does not solve the problem of instability of highly overmoded opto-mechanical system. Therefore, a universal method of PI suppression is still needed. 

We here propose a solution based on optimization of the shape of the cavity mirrors leading to increase of the diffraction losses of all high-order optical modes of the optical cavity and subsequent increase of the PI threshold. Note that diffraction losses a cavity modes can be increased rather significantly by properly shaping mirrors of the cavity \cite{92pare,05kuznetsov,07tiffani}, however, with loss increase of main mode which is inappropriate for laser gravitational detector detector. We propose the method allows realizing an optical cavity containing only one family of low loss bounded modes. This is achievable in the case of large area optical mirrors. In a realistic gravity wave detector, though, the size of the mirrors is limited and suppression of the high-order modes is associated with loss increase of the fundamental mode family. We study this practically interesting case using numerical simulations and show that it is feasible to increase the PI threshold at least by an order of magnitude by 
proper shaping mirrors of a LIGO interferometer keeping diffraction loss of main mode at acceptable low level. We show that the stability of the modified interferometer with respect to the mirror tilt and shape variations is acceptable. Finally, we found that the optimized cavity can still be interrogated using
conventional Gaussian beams.

\section{Threshold condition}

The lowest PI intracavity threshold power evaluated for a Fabry-Perot (FP) resonator can be found from expression \cite{01a1BrStVy}  
\begin{eqnarray}
\label{P}
P_{th}& =& \frac{m c^2\Omega_m^2 \mathcal L} {4 \zeta \omega_s Q_m},\\
\zeta &=& \frac{V \left|\int_{z=0} f_p (\vec r\bot)\, f_s (\vec r_\bot) \vec u_z(\vec r) d\vec r_\bot\right|^2}{
  \int |\vec f_p (\vec r_\bot)|^2  d\vec r_\bot \int |\vec f_s (\vec r_\bot)|^2  d\vec r_\bot
  \int |\vec u(\vec r)|^2 d V} \nonumber
\end{eqnarray}
where $\mathcal L$ is the round trip optical attenuation coefficient of the Stokes mode, $m$ is the mass of the mirror, or test mass, $Q_m$ is the quality factor of the elastic mode, $c$ is the speed of light in the vacuum, $\zeta$ is a mismatching factor, $V$ is volume of the mirror, $\vec u(\vec r)$ is the mechanical mode displacement, $u_z$ is the same normal displacement on  the mirror surface, and $f_p,\ f_s$ is distribution of main and Stokes optical modes on mirror surface. The integration is performed over mirror volume ($dV$) and mirror surface ($d\vec r_\bot$).

Equation (\ref{P}) is obtained for the all-resonant case: $\omega_p=\omega_s +\Omega_m$. Substituting to Eq.~(\ref{P}) parameters of LIGO system, presented in Table~(\ref{param}), and assuming full overlapping ($\zeta =1$), we find that the PI threshold power, $P_{th}$, is more than two orders of magnitude smaller if compared with the envisioned power level $P$ \cite{01a1BrStVy}. To increase the threshold towards the desirable value we propose to increase $\mathcal L$ to $ 8\,000\, \text{ppm}$ by inducing leakage of the Stokes light out of the cavity due to enhanced diffraction of the high order optical modes. This increase results in a small practically acceptable increase of the attenuation of the fundamental mode $\mathcal L_p=5\, \text{ppm}$.
\begin{table}[ht]
\caption{Parameters of LIGO used in calculations}\label{param}
\begin{tabular}{|l|c|}
\hline
Parameter & Value \\
\hline
Arm length, $L$ & $4$~km\\
Optical wavelength, $\lambda$ & $1064$~nm\\
Intracavity power, $P$ & $800$~kW\\
$AS_{00}$ (main Gaussian) mode round trip loss, ${\mathcal L}_p$ & 0.45~ppm\\
$D_{10}$ ($LG_{10}$) dipole mode round trip loss, ${\mathcal L}$ & 10~ppm\\
Characteristic cavity length $b=\sqrt{L\lambda/2\pi}$ & $0.0260$~m\\
Radius of mirrors, $R$ &$0.17$~m\\
Dimensionless mirror radius $a_m=R/b$ & $6.53$ \\
Radius $w$ of laser spot at the mirror for TEM00 mode & $0.06$~m \\
Radius $w_0$ of laser beam at the waist  & $ 0.0115$~m\\
Curvature radius of spherical mirrors, $R_c$ & $2076$~m\\
Geometric parameter $g=1-L/R_c$  of the cavity & $-0.92649$\\
Gouy phase, $\arctan \left[ (b/w_0)^2\right ]$ & 1.378\\
\hline
\end{tabular}
\end{table}

The idea of the method relies on a dependence of the attenuation of high order modes of a FP cavity on relatively small deviation of the cavity mirror shape from the spherical one. The diffraction loss of the fundamental axial symmetric mode decreases exponentially with, while the loss of the other modes follows a power law dependence on the mirror diameter in a properly designed cavity. The ratio of round trip losses of  the fundamental and the lowest loss higher order optical modes of a cavity should exceed two orders of magnitude. To compare, LIGO cavity has this ratio fixed at the level of 20 for the main and first dipole modes (Table~\ref{param}). As shown in the next section, minute modifications of the LIGO mirror shape, keeping the overall mirror size intact, results in a significant increase of round trip loss of unwanted optical modes and increase of the PI threshold towards desirable numbers.

\section{Mirror shape optimization}

We consider resonators having nearly Gaussian spatial profile of the lowest order modes to ensure that the conventional auxiliary optics can be utilized with them. This is important for the post-processing of the output light requiring perfect matching with the modes of conventional filtering cavities as well as local oscillators involved in the data acquisition. We found that this condition is fulfilled if the curvature of the mirrors stays the same as the curvature of spherical mirrors of the conventional cavity.

Below we use dimensionless variables and parameters:
\begin {align}
 x & = \frac{r}{b},\quad b=\sqrt\frac{L\lambda}{2\pi},\quad \rho= \frac{R_c}{L}\ , \quad a_m=\frac{r_m}{b}\,,
\end {align}
where $r$ is distance from centre of mirror, $b$ is scaling factor, $L$ is distance  between mirrors, $\lambda$ is a wavelength, $R_c$ is curvature radius of mirror, $r_m$ is radius of mirror.
The shape of the mirrors of the FP cavity is described by
\begin{equation}
 \label{profile}
 y =y_0 \left(1 - e^{-z -\alpha z^2 -\beta z^3 } \right), \quad z=\frac{x^2}{2\rho y_0}
\end{equation}
where   $\alpha$, $\beta$ and $y_0$ are dimensionless independent parameters we optimize. The profile (\ref{profile}) transforms into spherical one $y=x^2/2\rho$ at $y_0\rightarrow \infty$  (or at $x \rightarrow 0$).

\begin{figure}[h]
	\includegraphics[width=0.45\textwidth]{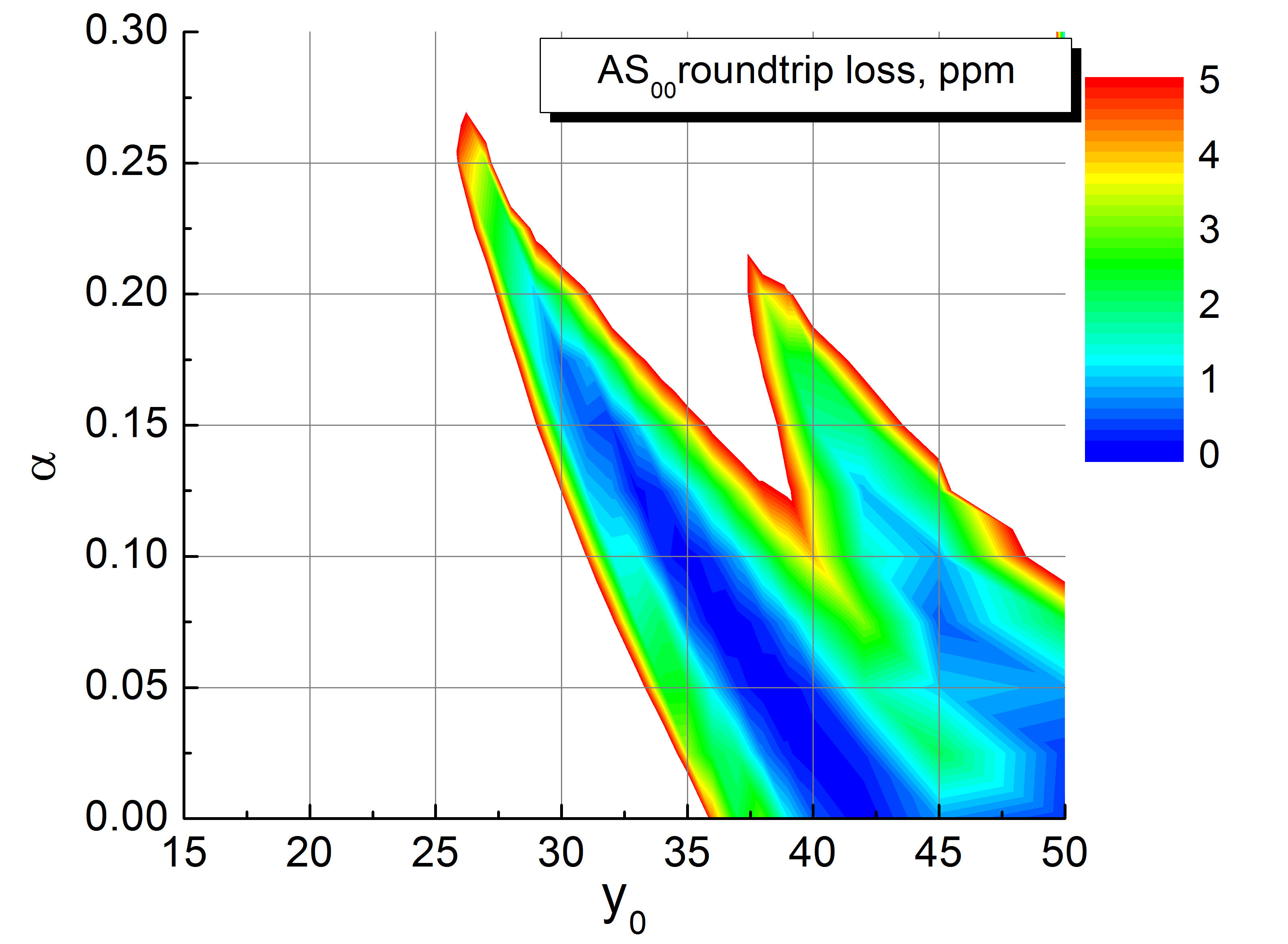}
	\caption{Dependence $AS_{00}$ mode round trip loss (ppm) on mirror shape parameters in range $y_0 = 15 \div 50 $ and $\alpha = 0 \div 0.3$. }	\label{loss}
\end{figure}
\begin{figure}
	\includegraphics[width=0.45\textwidth]{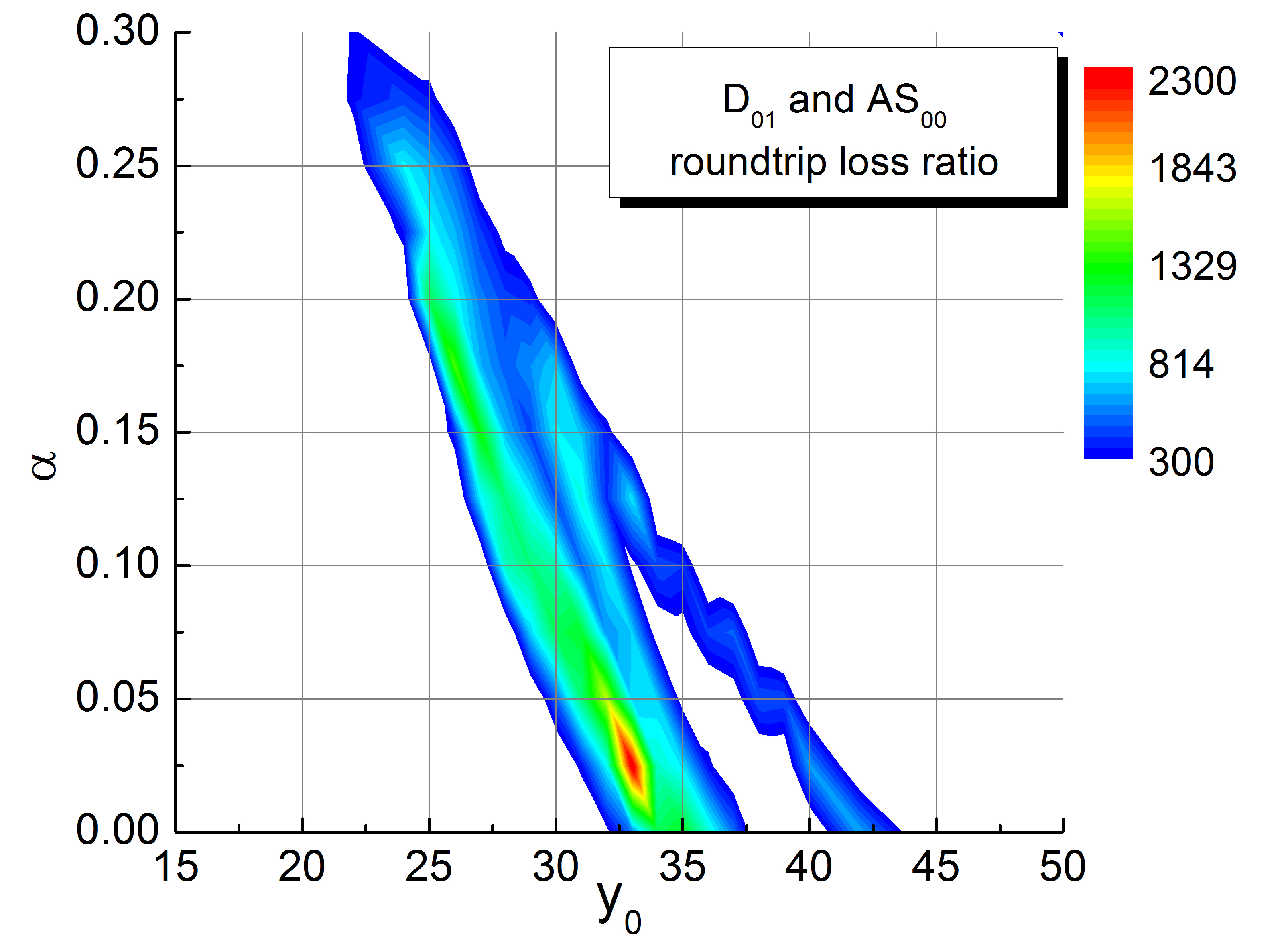}
	\caption{Dependence attenuation ratio $D_{10}$ to $AS_{00}$ on mirror shape parameters in range $y_0 = 15 \div 50 $ and $\alpha = 0 \div 0.3$. }	\label{distribution}
\end{figure}

While fundamental understanding of the optimization procedure can be gained from Born-Oppenheimer approach applied to a FP cavity, an accurate analytical optimization of the mirror shape is unfeasible, so numerical simulations are to be used. We utilize a  matrix analogue of Fresnel integral  to find electric field distribution $\Psi^\text{right}$ on the right mirror of the FP resonator using distribution $\Psi^\text{left}$ on the left mirror (and vice versa): $\Psi^\text{right} = RPR\, \Psi^\text{left}$, where matrix $P$ describes propagation from left plane to right one and depends on the mode of the cavity, and diagonal matrix  $R$ describes shapes of the mirrors. Equation $ (RPR)^2\, \Psi = \Lambda \Psi$ is solved numerically and round trip loss is found from $\mathcal L=1-|\Lambda|^2$.

Following the approach described in \cite{14PRAa1fdvmm, 93vinet} we define propagation matrix for axial symmetric (AS) modes through Hankel transform as
\begin{align}
\label{HT1}
P &= H^+ G (H^+)^{-1}, \quad
G_{kn} =\exp\left(-i\frac{\xi_k^2}{2a^2}\right)  \delta_{nk}, \\
\label{HT2}
H^+_{kn}&=\frac{4\pi a^2}{\xi_N^2 N_n}\,J_0\left(\frac{\xi_k\xi_n}{\xi_N}\right), \
N_n =\left[1+\frac{1}{\xi_n^2}\right] J_0(\xi_n^2),
\end{align}
where $a=Sa_m$, $S\simeq 1.5\cdots 5>1$ is window parameter, $J_0$ and $J_1$ are Bessel functions of the first kind, $\xi_n$ is set of first $N$ roots of characteristic equation $J_0(\xi) -\xi J_1(\xi)=0$.

The propagator $P$ for azimuthal (not axial symmetric) higher order modes is easy to generalize, for example, for dipole modes (dependence on azimuthal angle $\phi$ as $\sim e^{i\phi}$) we have to substitute $J_1$ instead $J_0$ in \eqref{HT2} and to use roots of  characteristic equation $J_1(\xi) -\xi J_2(\xi)=0$.

The mirror shape is presented numerically by matrix
 \begin{align}
 \label{R}
  R_{kn} &= d_k e^{-iy(x_n)}\delta_{kn}, \quad
  x_n = \xi_n a/\xi_N, \\
	d_k &= \left\{
	\begin{array}{ll}
	 1, \quad &\text{if}\ x_n<a_m,\\
	 0 \quad &\text{if} \ x_n>a_m
	\end{array}
	 \right.
 \end{align}
where coefficients $d_k$  define reflective surface of mirror.

\begin{figure}
	\includegraphics[width=0.45\textwidth]{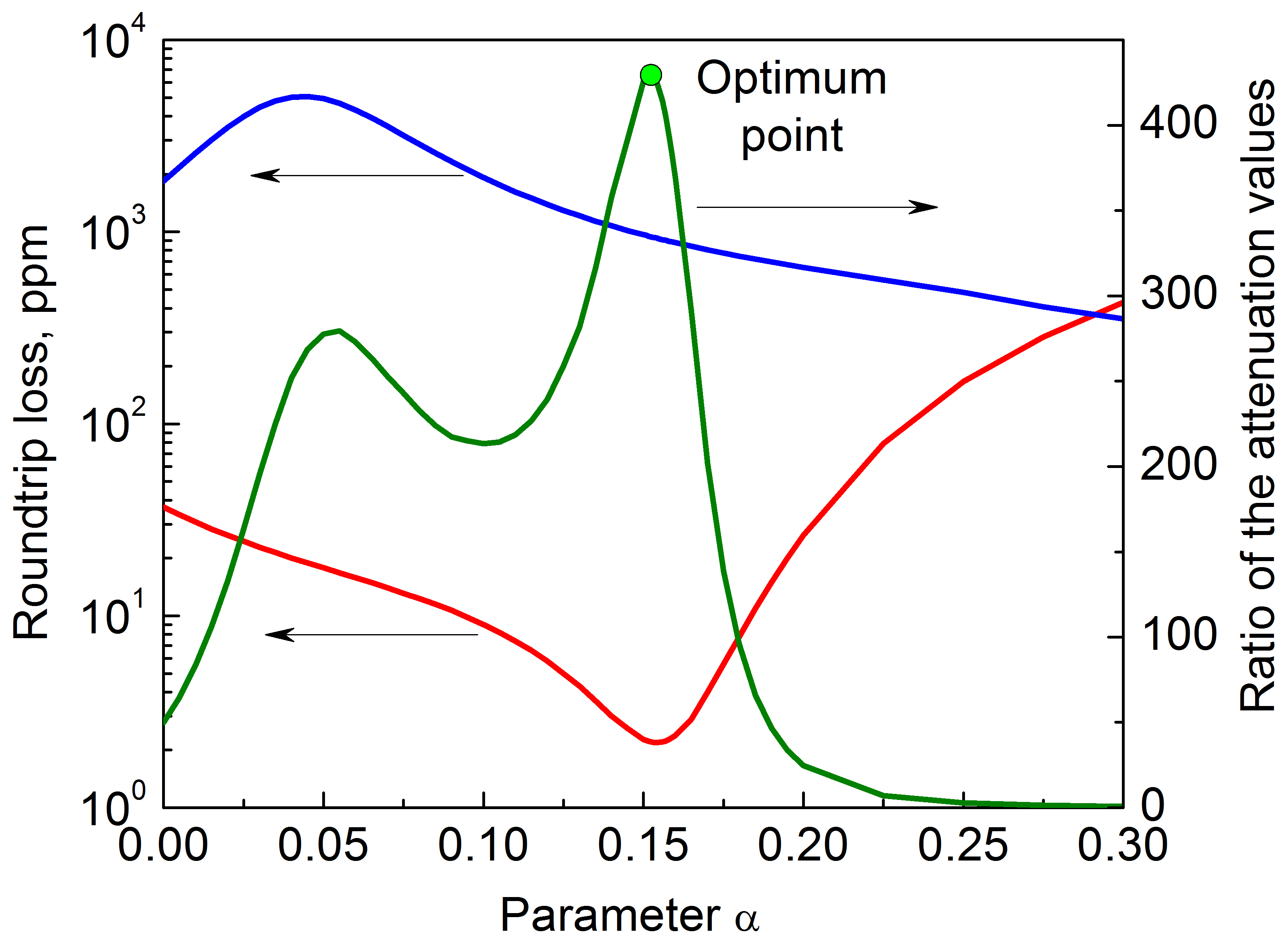}
	\caption{Searching of local optimum point under following conditions: $AS_{00}$ round trip loss (red curve) not exceeds 5 ppm, $D_{10}$ round trip loss (blue curve) is approximately $10^3$ ppm and ratio of these losses (green curve) reach a local maximum. This figure corresponds to the $1$ parameters set in Table~(\ref{table1}). }\label{opt_loss}
\end{figure}

Selecting point number $N=512$ and window parameter $S=2$ we found dependence of the attenuation parameters of various modes of the resonator on the mirror shapes (Fig.~\ref{loss}). As the rule, the first dipole mode ($D_{10}$) has the lowest loss with respect to the fundamental axial symmetric mode ($AS_{00}$). We optimized the problem by identifying local maxima of the ratio of attenuation of the dipole and the fundamental mode, as illustrated by Figs.~\ref{distribution}, \ref{opt_loss}. Several identified local optima for the mirror shape are listed in Table~(\ref{table1}).
We choose the radius of curvature  in the center of the deformed mirrors to be $R_c=2014$~m  (it corresponds to spot radius $w=0.09$~m for spherical mirror). 
The simulation shows that modification of the mirror shape results in significant increase of diffraction loss of the high order modes while keeping the attenuation of the fundamental modes to be low.

\begin{figure}
 \includegraphics[width=0.45\textwidth]{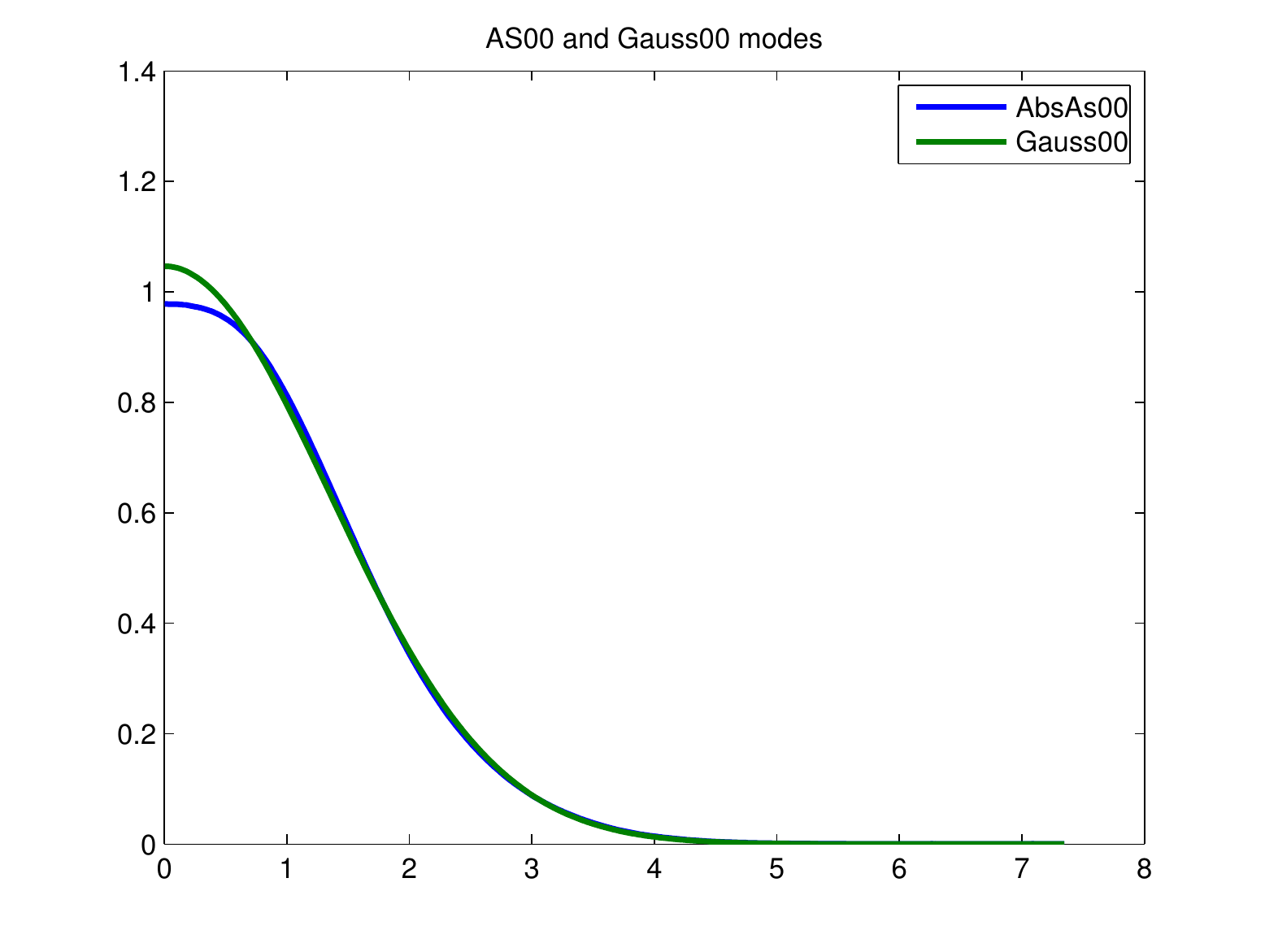}
 \caption{Amplitude distribution of main AS mode corresponding to parameters set 3 in Table~\ref{table1} and Gaussian one.}\label{ASgauss}
\end{figure}

The amplitude distribution of the modes of interest only slightly differs from the Gaussian fit having the same full width at the half maximum (corresponding to spot radius about $0.05$~m) --- see Fig.~\ref{ASgauss}. Normalizing the electric field amplitude of the modes over the beam cross section as $\int |E|^2 dS=1$ we find that the mismatch between the Gaussian fit and the cavity eigenmode determined as   $\int\left(|E_{AS00}|-E_\text{Gauss}\right)^2 dS<10^{-3}$ for any of the selected mirror shape (here we put $\left|E_{AS00}\right|$ because numerically found eigen mode has imagine part, but it is relatively small $< 10^{-4}$). It means that the resonator with optimized mirrors can be pumped using Gaussian beams and the quantum state of the light exiting the resonator can be analysed using Gaussian shaped local oscillator beam.

\begin{widetext}
\begin{center}
 \begin{table}[h]
   \caption{Values of the round trip loss (ppm) for FP cavities having spherical and deformed mirrors, calculated numerically with points number $N=512$ and window parameter $S=2$. We used LIGO parameters summarized in Table~\ref{param} for the FP with spherical mirrors according to laser spot radius of $w=0.06$~m on mirror. $AS$, $D$, $Q$, and $M$, stand for the axial symmetric, dipole, quadrupole, and hexapole modes. }\label{table1}
    \begin{tabular}{||c|c|c|c|c|c|c|c|c|c|c||}
    \hline \hline
  & Modes   &$AS_{00}$ &  $AS_{01}$ & $AS_{02}$ & $D_{10}$ & $D_{11}$ & $Q_{20}$ & $Q_{21}$ & $M_{30}$& $M_{31}$\\
      \hline\hline
   & Spherical &0.45 & 170  & 6500 & 8.9 & 1050 & 100 & 5100 & 470 & 20\,000 \\
   \hline
   $1$&  $y_0 =20,\, \alpha=0.1525,\ \beta=0.35$& 2.2 & 46\,000 & 43\,000 & 940 & 20\,000 & 19\,000 & 30\,000 & 10\,000 & 28\,000 \\
    $2$&  $y_0 =27.5,\, \alpha=0.21,\ \beta=0$ & 2.6 & 46\,000 & 19\,000 & 1100 & 41\,000 & 23\,000  & 16\,000 & 11\,000 & 30\,000\\
  3&  $y_0 =30,\, \alpha=0.175,\ \beta=-0.05$ & 3.3 & 37\,000 & 20\,000 & 1600 & 36\,000 & 19\,000 & 17\,000 & 8800 & 12\,000\\
    \hline \hline
    \end{tabular}
    \end{table}
    \end{center}
\end{widetext}

The tolerance requirements to approach the parameters described in Table~(\ref{table1}) are reasonable. For example, for variant 3 in Table~\ref{table1} the parameters $y_0$, $\alpha$, and $\beta$ have to hold with accuracy  about $\pm 0.25$, $\pm 0.005$, and $\pm 0.005$ respectively to keep the value of the loss within 3~dB of the predicted values. It means that shape of the mirrors has to be manufactured within tolerance $\pm 0.03\, \lambda$, which is practically feasible. Another important factor is related to the dynamic stability of the modified FP cavity, which is discussed in the next section.

\section{Tilt stability}

The optimization of the mirrors shape results in reduction of the radiative loss of the higher order FP modes. As the consequence, the resonator sensitivity to the tilt of the mirrors increases if compared with the sensitivity of a conventional FP resonator. The tilt lifts orthogonality and results in linear coupling among the optical modes. The coupling is largest for the axial symmetric and the dipole modes. It is reasonable to expect that the angle sensitivity of the cavity attenuation is approximately proportional to the square root of the clipping loss value of the dipole mode.

There is no known way of accurate analytical evaluation of the loss increase due to mirror tilt. Moreover, the numerical simulations become rather involved since the tilt breaks the symmetry of the system. To evaluate this effect we use method of successive approximations that is based on fusion of the both numerical and analytical methods. According to this method, the round trip loss depends on small tilt angle $\theta$ of one of the mirrors of the FP cavity as
\begin{align}
\label{LsphN}
 \tilde {\mathcal L}_{00} &\simeq  \mathcal L_{00} \left\{1 +   \frac{\theta^2}{\theta_\text{perm}^2}\right\},\quad
      \frac{1}{\theta_\text{perm}^2}= \frac{kL S_U}{\mathcal L_{00}},\\
 S_U & \equiv \Re\left[U_{00,00} -  2\sum_{j}\frac{ \Lambda_{j}^2 \big|U_{j,00}\big|^2}{\Lambda_{j}^2 -\Lambda_{00}^2 }\right],\\ \nonumber
	U_{00,00} &\equiv   \int |\psi_{00}(x)|^2\, x^3\, dx,
	\quad  \int |\psi_{j }(x)|^2x\, dx=1\,,\\
	U_{j,00} &\equiv   \int \psi^*_{j }(x)\psi_{00 }(x)\, x^2\, dx\,.
\end{align}
where $k$ is the wave number, $\Lambda_{j}$ and $\psi_{j }$ are the calculated numerically forward trip eigenvalue and eigenvector of the unperturbed problem (no tilt), $\theta_\text{perm}$ is a permissible angle to characterize tilt stability.

Numeric calculations for parameters sets 1,2,3 listed in Table~\ref{table1} give  the following  permissible tilt angles:
 \begin{align}
  \theta_\text{perm}^{(1)}&\simeq 0.12\ \mu\text{rad},\quad
      \theta_\text{perm}^{(2)}\simeq \theta_\text{perm}^{(3)}\simeq 0.08\ \mu\text{rad}
 \end{align}
To figure out if this value is large, we calculate similar number for current LIGO interferometer (Table~\ref{param}) and find $\theta_{LIGO} \simeq 0.6\ \mu$rad. In other words, the dynamic range of the mirror angle deviation of conventional interferometer is an order of magnitude better than one of the interferometer with modified mirrors. This is expected as the loss parameter of the first dipole mode is approximately 1,000~ppm (10~ppm) for the single mode (conventional) resonator.

The considered here model of a FP resonator is a simplification of the realistic LIGO system with one exception. The last one includes two FP resonators and a recycling mirror that allows increasing the effective finesse of the multi-resonator system. It is possible to show that the LIGO resonator effective loss is proportional to $T_1T_2/4+{ \mathcal L_1}$, where $T_1$ and $T_2$ are power transmission coefficients of the input and recycling mirrors, respectively, and ${\mathcal L}_1$ is the attenuation per round trip in the FP resonator. In advanced LIGO interferometer $T_1=0.014$ and $T_2=0.03$ \cite{aLIGO2014b}, so the effective transmission coefficient is about $T_\text{eff}=T_1T_2/4=100$~ppm, whereas diffraction loss ${ \mathcal L}_1\simeq 0.45$~ppm. Hence, increase  of diffraction loss of the fundamental mode by about an order of magnitude (to $5$~ppm) seems to be acceptable to use squeezing about $\mathcal L_{00}/T_\text{eff}\simeq 0.02$. The increase of the attenuation of the dipole mode beyond 
100~ppm 
results in reduction of PI in 
accordance with Eq.~\ref{P}.

\section{Conclusion}

We have shown that one can reduce impact of parametric instability of an interferometric gravity wave detector by optimizing geometrical shape of its mirrors. The improvement stems from the dependence of the threshold of the instability on the losses of the optical modes involved in the process. Modification of the mirror shape increases the diffraction loss of the higher order optical modes resulting in the instability threshold increase, occurring at the cost of scrutinizing the mirror tilt stability requirements. To explain the effect, we have created a semi-analytical model of the diffraction loss of a Fabry-Perot cavity having an arbitrary mirror shape and found that engineered increase of diffraction loss of the high order modes with necessity leads to larger sensitivity of the fundamental mode loss to the mirror tilt. Optimizing the ratio of the losses of the cavity modes it is possible to achieve a significant suppression of the opto-mechanic 
instability and also keep acceptable tolerances of the system implementation. We validated results of our predictions with numerical simulations.

\section{Acknowledgments}

The authors acknowledge fruitful discussions with William Kells. Mikhail V. Poplavskiy and Sergey P. Vyatchanin acknowledge support from the Russian Foundation for Basic Research (Grant No. 14-02-00399A and Grant No. 13-02-92441 in frame of program ASPERA), and National Science Foundation (Grant No. PHY-130586).


\begin{thebibliography}{22}
\expandafter\ifx\csname natexlab\endcsname\relax\def\natexlab#1{#1}\fi
\expandafter\ifx\csname bibnamefont\endcsname\relax
  \def\bibnamefont#1{#1}\fi
\expandafter\ifx\csname bibfnamefont\endcsname\relax
  \def\bibfnamefont#1{#1}\fi
\expandafter\ifx\csname citenamefont\endcsname\relax
  \def\citenamefont#1{#1}\fi
\expandafter\ifx\csname url\endcsname\relax
  \def\url#1{\texttt{#1}}\fi
\expandafter\ifx\csname urlprefix\endcsname\relax\def\urlprefix{URL }\fi
\providecommand{\bibinfo}[2]{#2}
\providecommand{\eprint}[2][]{\url{#2}}

\bibitem[{\citenamefont{LVC-Collaboration}(2013)}]{aLIGO2013}
\bibinfo{author}{\bibnamefont{LVC-Collaboration}}, \bibinfo{journal}{arXiv}
  \textbf{\bibinfo{volume}{1304.0670}} (\bibinfo{year}{2013}).

\bibitem[{\citenamefont{Dooley et~al.}(2014)\citenamefont{Dooley, Akutsu,
  Dwyer, and Puppo}}]{aLIGO2014}
\bibinfo{author}{\bibfnamefont{K.}~\bibnamefont{Dooley}},
  \bibinfo{author}{\bibfnamefont{T.}~\bibnamefont{Akutsu}},
  \bibinfo{author}{\bibfnamefont{S.}~\bibnamefont{Dwyer}}, \bibnamefont{and}
  \bibinfo{author}{\bibfnamefont{P.}~\bibnamefont{Puppo}},
  \bibinfo{journal}{arXiv} \textbf{\bibinfo{volume}{1411.6068}}
  (\bibinfo{year}{2014}).

\bibitem[{\citenamefont{{V.B.Braginsky S.E.Strigin and
  S.P.Vyatchanin}}(2001)}]{01a1BrStVy}
\bibinfo{author}{\bibnamefont{{V.B.Braginsky S.E.Strigin and S.P.Vyatchanin}}},
  \bibinfo{journal}{Physics Letters A} \textbf{\bibinfo{volume}{287}},
  \bibinfo{pages}{331} (\bibinfo{year}{2001}).

\bibitem[{\citenamefont{{V. B. Braginsky, S. E. Strigin and
  S.P.Vyatchanin}}(2002)}]{02a1BrStVy}
\bibinfo{author}{\bibnamefont{{V. B. Braginsky, S. E. Strigin and
  S.P.Vyatchanin}}}, \bibinfo{journal}{Physics Letters A}
  \textbf{\bibinfo{volume}{305}}, \bibinfo{pages}{111} (\bibinfo{year}{2002}).

\bibitem[{\citenamefont{Chen et~al.}(2014)\citenamefont{Chen, C.Zhao,
  Danilishin, L.~Ju, Wang, Vyatchanin, Molinelli, Kuhn, Gras, Briant
  et~al.}}]{14ArxivChen}
\bibinfo{author}{\bibfnamefont{X.}~\bibnamefont{Chen}},
  \bibinfo{author}{\bibnamefont{C.Zhao}},
  \bibinfo{author}{\bibfnamefont{S.}~\bibnamefont{Danilishin}},
  \bibinfo{author}{\bibfnamefont{D.~B.} \bibnamefont{L.~Ju}},
  \bibinfo{author}{\bibfnamefont{H.}~\bibnamefont{Wang}},
  \bibinfo{author}{\bibfnamefont{S.~P.} \bibnamefont{Vyatchanin}},
  \bibinfo{author}{\bibfnamefont{C.}~\bibnamefont{Molinelli}},
  \bibinfo{author}{\bibfnamefont{A.}~\bibnamefont{Kuhn}},
  \bibinfo{author}{\bibfnamefont{S.}~\bibnamefont{Gras}},
  \bibinfo{author}{\bibfnamefont{T.}~\bibnamefont{Briant}},
  \bibnamefont{et~al.}, \bibinfo{journal}{arXiv}
  \textbf{\bibinfo{volume}{1411.3016}} (\bibinfo{year}{2014}).

\bibitem[{\citenamefont{Kippenberg et~al.}(2005)\citenamefont{Kippenberg,
  Rokhsari, Carmon, Scherer, and Vahala}}]{05Kippenberg}
\bibinfo{author}{\bibfnamefont{T.~J.} \bibnamefont{Kippenberg}},
  \bibinfo{author}{\bibfnamefont{H.}~\bibnamefont{Rokhsari}},
  \bibinfo{author}{\bibfnamefont{T.}~\bibnamefont{Carmon}},
  \bibinfo{author}{\bibfnamefont{A.}~\bibnamefont{Scherer}}, \bibnamefont{and}
  \bibinfo{author}{\bibfnamefont{K.~J.} \bibnamefont{Vahala}},
  \bibinfo{journal}{Phys. Rev. Lett} \textbf{\bibinfo{volume}{95}},
  \bibinfo{pages}{033901} (\bibinfo{year}{2005}).

\bibitem[{\citenamefont{Matsko et~al.}(2009)\citenamefont{Matsko, Savchenkov,
  Ilchenko, Seidel, and Maleki}}]{09PRLmatsko}
\bibinfo{author}{\bibfnamefont{A.~B.} \bibnamefont{Matsko}},
  \bibinfo{author}{\bibfnamefont{A.~A.} \bibnamefont{Savchenkov}},
  \bibinfo{author}{\bibfnamefont{V.~S.} \bibnamefont{Ilchenko}},
  \bibinfo{author}{\bibfnamefont{D.}~\bibnamefont{Seidel}}, \bibnamefont{and}
  \bibinfo{author}{\bibfnamefont{L.}~\bibnamefont{Maleki}},
  \bibinfo{journal}{Phys. Rev. Lett} \textbf{\bibinfo{volume}{103}},
  \bibinfo{pages}{257403} (\bibinfo{year}{2009}).

\bibitem[{\citenamefont{Matsko et~al.}(2012)\citenamefont{Matsko, Savchenkov,
  and Maleki}}]{12OEmatsko}
\bibinfo{author}{\bibfnamefont{A.~B.} \bibnamefont{Matsko}},
  \bibinfo{author}{\bibfnamefont{A.~A.} \bibnamefont{Savchenkov}},
  \bibnamefont{and} \bibinfo{author}{\bibfnamefont{L.}~\bibnamefont{Maleki}},
  \bibinfo{journal}{Opt. Express} \textbf{\bibinfo{volume}{20}},
  \bibinfo{pages}{16234} (\bibinfo{year}{2012}).

\bibitem[{\citenamefont{Evans et~al.}(2015)\citenamefont{Evans, Gras,
  P.Fritschel, and et~al}}]{15ArxivEvans}
\bibinfo{author}{\bibfnamefont{M.}~\bibnamefont{Evans}},
  \bibinfo{author}{\bibfnamefont{S.}~\bibnamefont{Gras}},
  \bibinfo{author}{\bibnamefont{P.Fritschel}}, \bibnamefont{and}
  \bibinfo{author}{\bibnamefont{et~al}}, \bibinfo{journal}{Physical Review
  Letters} \textbf{\bibinfo{volume}{114}} (\bibinfo{year}{2015}), \eprint{arXiv
  1502.06058}.

\bibitem[{\citenamefont{C.Zhao et~al.}(2005)\citenamefont{C.Zhao, L.Ju,
  J.Degallaix, S.Gras, and D.G.Blair}}]{05PRLzhao}
\bibinfo{author}{\bibnamefont{C.Zhao}}, \bibinfo{author}{\bibnamefont{L.Ju}},
  \bibinfo{author}{\bibnamefont{J.Degallaix}},
  \bibinfo{author}{\bibnamefont{S.Gras}}, \bibnamefont{and}
  \bibinfo{author}{\bibnamefont{D.G.Blair}}, \bibinfo{journal}{Physical Review
  Letters} \textbf{\bibinfo{volume}{94}}, \bibinfo{pages}{121102}
  (\bibinfo{year}{2005}).

\bibitem[{\citenamefont{S.Gras et~al.}(2010)\citenamefont{S.Gras, C.Zhao,
  D.G.Blair, and L.Ju}}]{10CQGgrass}
\bibinfo{author}{\bibnamefont{S.Gras}}, \bibinfo{author}{\bibnamefont{C.Zhao}},
  \bibinfo{author}{\bibnamefont{D.G.Blair}}, \bibnamefont{and}
  \bibinfo{author}{\bibnamefont{L.Ju}}, \bibinfo{journal}{Class. Quantum Grav.}
  \textbf{\bibinfo{volume}{27}}, \bibinfo{pages}{205019}
  (\bibinfo{year}{2010}).

\bibitem[{\citenamefont{J.Degallaix et~al.}(2007)\citenamefont{J.Degallaix,
  C.Zhao, L.Ju, and D.G.Blair}}]{07JOSABdegallaix}
\bibinfo{author}{\bibnamefont{J.Degallaix}},
  \bibinfo{author}{\bibnamefont{C.Zhao}}, \bibinfo{author}{\bibnamefont{L.Ju}},
  \bibnamefont{and} \bibinfo{author}{\bibnamefont{D.G.Blair}},
  \bibinfo{journal}{J. Opt. Soc. Am. B} \textbf{\bibinfo{volume}{24}},
  \bibinfo{pages}{1336} (\bibinfo{year}{2007}).

\bibitem[{\citenamefont{S.Gras et~al.}(2009)\citenamefont{S.Gras, D.G.Blair,
  and C.Zhao}}]{09GQGgras}
\bibinfo{author}{\bibnamefont{S.Gras}},
  \bibinfo{author}{\bibnamefont{D.G.Blair}}, \bibnamefont{and}
  \bibinfo{author}{\bibnamefont{C.Zhao}}, \bibinfo{journal}{Classical Quantum
  Gravity} \textbf{\bibinfo{volume}{26}}, \bibinfo{pages}{135012}
  (\bibinfo{year}{2009}).

\bibitem[{\citenamefont{Ju et~al.}(2009)\citenamefont{Ju, Blair, Zhao, Gras,
  Zhang, Barriga, Miao, Fan, and Merrill}}]{09CQGju}
\bibinfo{author}{\bibfnamefont{L.}~\bibnamefont{Ju}},
  \bibinfo{author}{\bibfnamefont{D.~G.} \bibnamefont{Blair}},
  \bibinfo{author}{\bibfnamefont{C.}~\bibnamefont{Zhao}},
  \bibinfo{author}{\bibfnamefont{S.}~\bibnamefont{Gras}},
  \bibinfo{author}{\bibfnamefont{Z.}~\bibnamefont{Zhang}},
  \bibinfo{author}{\bibfnamefont{P.}~\bibnamefont{Barriga}},
  \bibinfo{author}{\bibfnamefont{H.}~\bibnamefont{Miao}},
  \bibinfo{author}{\bibfnamefont{Y.}~\bibnamefont{Fan}}, \bibnamefont{and}
  \bibinfo{author}{\bibfnamefont{L.}~\bibnamefont{Merrill}},
  \bibinfo{journal}{Classical Quantum Gravity} \textbf{\bibinfo{volume}{26}},
  \bibinfo{pages}{015002} (\bibinfo{year}{2009}).

\bibitem[{\citenamefont{{V. B. Braginsky and S.P.Vyatchanin}}(2002)}]{02a1BrVy}
\bibinfo{author}{\bibnamefont{{V. B. Braginsky and S.P.Vyatchanin}}},
  \bibinfo{journal}{Physics Letters A} \textbf{\bibinfo{volume}{293}},
  \bibinfo{pages}{228} (\bibinfo{year}{2002}).

\bibitem[{\citenamefont{Miller et~al.}(2011)\citenamefont{Miller, Evans,
  Barsotti, Fritschel, MacInnis, Mittleman, Shapiro, Soto, and
  Torrie}}]{11PRAmiller}
\bibinfo{author}{\bibfnamefont{J.}~\bibnamefont{Miller}},
  \bibinfo{author}{\bibfnamefont{M.}~\bibnamefont{Evans}},
  \bibinfo{author}{\bibfnamefont{L.}~\bibnamefont{Barsotti}},
  \bibinfo{author}{\bibfnamefont{P.}~\bibnamefont{Fritschel}},
  \bibinfo{author}{\bibfnamefont{M.}~\bibnamefont{MacInnis}},
  \bibinfo{author}{\bibfnamefont{R.}~\bibnamefont{Mittleman}},
  \bibinfo{author}{\bibfnamefont{B.}~\bibnamefont{Shapiro}},
  \bibinfo{author}{\bibfnamefont{J.}~\bibnamefont{Soto}}, \bibnamefont{and}
  \bibinfo{author}{\bibfnamefont{C.}~\bibnamefont{Torrie}},
  \bibinfo{journal}{Physics Letters A} \textbf{\bibinfo{volume}{375}},
  \bibinfo{pages}{788} (\bibinfo{year}{2011}).

\bibitem[{\citenamefont{Pare et~al.}(1992)\citenamefont{Pare, Gagnon, and
  Belanger}}]{92pare}
\bibinfo{author}{\bibfnamefont{C.}~\bibnamefont{Pare}},
  \bibinfo{author}{\bibfnamefont{L.}~\bibnamefont{Gagnon}}, \bibnamefont{and}
  \bibinfo{author}{\bibfnamefont{P.~A.} \bibnamefont{Belanger}},
  \bibinfo{journal}{Phys. Rev. A} \textbf{\bibinfo{volume}{46}},
  \bibinfo{pages}{4150} (\bibinfo{year}{1992}).

\bibitem[{\citenamefont{Kuznetsov et~al.}(2005)\citenamefont{Kuznetsov, Stern,
  and Coppeta}}]{05kuznetsov}
\bibinfo{author}{\bibfnamefont{M.}~\bibnamefont{Kuznetsov}},
  \bibinfo{author}{\bibfnamefont{M.}~\bibnamefont{Stern}}, \bibnamefont{and}
  \bibinfo{author}{\bibfnamefont{J.}~\bibnamefont{Coppeta}},
  \bibinfo{journal}{Opt. Express} \textbf{\bibinfo{volume}{13}},
  \bibinfo{pages}{171} (\bibinfo{year}{2005}).

\bibitem[{\citenamefont{Tiffany and Leger}(2007)}]{07tiffani}
\bibinfo{author}{\bibfnamefont{B.}~\bibnamefont{Tiffany}} \bibnamefont{and}
  \bibinfo{author}{\bibfnamefont{J.}~\bibnamefont{Leger}},
  \bibinfo{journal}{Opt. Express} \textbf{\bibinfo{volume}{15}},
  \bibinfo{pages}{13463} (\bibinfo{year}{2007}).

\bibitem[{\citenamefont{Ferdous et~al.}(2014)\citenamefont{Ferdous, Demchenko,
  Vyatchanin, Matsko, and Maleki}}]{14PRAa1fdvmm}
\bibinfo{author}{\bibfnamefont{F.}~\bibnamefont{Ferdous}},
  \bibinfo{author}{\bibfnamefont{A.}~\bibnamefont{Demchenko}},
  \bibinfo{author}{\bibfnamefont{S.}~\bibnamefont{Vyatchanin}},
  \bibinfo{author}{\bibfnamefont{A.}~\bibnamefont{Matsko}}, \bibnamefont{and}
  \bibinfo{author}{\bibfnamefont{L.}~\bibnamefont{Maleki}},
  \bibinfo{journal}{Phys. Rev A} \textbf{\bibinfo{volume}{90}},
  \bibinfo{pages}{033826} (\bibinfo{year}{2014}).

\bibitem[{\citenamefont{Vinet and Hello}(1993)}]{93vinet}
\bibinfo{author}{\bibfnamefont{J.}~\bibnamefont{Vinet}} \bibnamefont{and}
  \bibinfo{author}{\bibfnamefont{P.}~\bibnamefont{Hello}},
  \bibinfo{journal}{Journal of Modern Optics} \textbf{\bibinfo{volume}{40}},
  \bibinfo{pages}{1981} (\bibinfo{year}{1993}).

\bibitem[{\citenamefont{LIGO-Collaboration}(2014)}]{aLIGO2014b}
\bibinfo{author}{\bibnamefont{LIGO-Collaboration}}, \bibinfo{journal}{arXiv}
  \textbf{\bibinfo{volume}{1411.4547}} (\bibinfo{year}{2014}).

\end{thebibliography}
\end{document}